\documentclass[prd,twocolumn,eqsecnum,nofootinbib]{revtex4}
\usepackage{bm}
\usepackage{amssymb}
\usepackage{graphicx}
\usepackage{epstopdf}
\usepackage{mathrsfs}
\usepackage{amsmath}

\begin{document}
\title{Physical objects approaching the Cauchy horizon of a rapidly rotating Kerr black hole} 
\author{Caroline Mallary$^1$, Gaurav Khanna$^1$ and Lior M.~Burko$^2$}
\affiliation{
$^1$ Department of Physics, University of Massachusetts, Dartmouth, Massachusetts  02747 \\
$^2$ School of Science and Technology, Georgia Gwinnett College, Lawrenceville, Georgia 30043}
\date{July 17, 2018}
\begin{abstract} 
We solve the 2+1-dimensional Teukolsky equation numerically for the Weyl scalars $\psi_0$ and $\psi_4$ along a time-like geodesic approaching the Cauchy horizon of a rapidly rotating perturbed Kerr black hole. We find that both the amplitude and frequency of the Weyl scalars agree with the results of linear perturbation analysis. We then model a physical object by a simple damped harmonic oscillator, which is driven by an external force that mimics the tidal force experienced by the infalling object. We use this model to find the total deformation of the object at the Cauchy horizon, and the resonant effect when the driving force's frequency matches the internal frequency of the oscillator that models the object. 
\end{abstract}
\maketitle

\section{Introduction }

The Strong Cosmic Censorship Conjecture (SCC) states that given generic compact or asymptotically-flat Cauchy data it is impossible  to extend the maximal Cauchy development with a Lorentzian manifold equipped with locally square-integrable connections \cite{Christodoulou-99}. This formulation of the SCC implies the generic spacetimes will always be globally hyperbolic, and that general relativity retains its deterministic predictability. 
This definition of the SCC is particularly useful when one seeks to find (possibly weak) solutions of the classical Einstein equations. Indeed, it is hard to see how one would make sense of the Einstein equations if this condition is not satisfied. 

However, the Einstein equations break down at singularities, when quantum gravity is expected to take over. That is, one should not be surprised that the classical Einstein equations no longer make sense when curvature becomes super-Planckian, even when the metric is $C^0$ extendible, as is the case with the mass-inflation singularity \cite{Ori-91,Ori-92}. 
From the point of view of an infalling observer, it therefore makes sense to ask whether an observer made of reasonable matter will detect the approach to a spacetime singularity, i.e., whether she will necessarily be destroyed upon approach to the Cauchy horizon (CH). A necessary condition for safe passage of the CH is that the answer to the latter question is in the negative. A definitive answer, however, must await the development of a quantum theory of gravity. In this paper we are interested in investigating this necessary condition for reasonable matter approaching the CH of perturbed isolated Kerr black holes. The rationale of this approach is that if extended physical objects cease to exist, it would be  (classically) because of very strong tidal deformations, and while the understanding of how many derivatives of the metric are continuous is extremely important for the mathematical investigation of black hole singularities, it is eventually the physical interactions that determine the infalling objects' fate.\footnote{One often considers unbounded tidal deformations as the criterion for ultimate destruction. Notice, however, that every physical material is effectively destroyed when experiencing tidal deformations that exceed a certain upper bound it can tolerate. When tidal deformations are unbounded, any physical object is ultimately destroyed. But if tidal deformations are bounded - even if very large - it becomes a practical engineering problem for materials scientists whether a material that can survive such deformations can be created, not a problem of principle. In other words, we are interested in the question of what the laws of physics forbid, not in what current engineering technology allows.}

The singularity inside black holes had long been thought to be spacelike (like the singularity inside an unperturbed Schwarzschild black hole), and possibly of the  Belinskii-Khalatnikov-Lifshitz (BKL) type \cite{BKL}. Poisson and Israel found that for an isolated spherical charged black hole that is perturbed by its own evolutionary collapse (i.e., by Price tails \cite{price}), a null singularity evolves at the Cauchy horizon \cite{poisson-israel}.  
The interaction of an infalling observer with the fields in the interior of a perturbed, isolated, spinning black hole was first considered by Ori \cite{Ori-91}, who argued, based on the metric and fields being  $C^0$ at the CH, that extended objects might not be destroyed, in the context of a spherical charged toy model perturbed by Price tail radiation. Burko and Ori \cite{Burko-Ori-95} considered the same model in greater detail, and modeled an infalling object as a (classical or quantum mechanical) undamped oscillator, and then solved for the deviation of the oscillator from its equilibrium length upon arrival at the CH (see also \cite{Burko-97a}). In \cite{Burko-Ori-95} it was found that such deviations are not just bounded, but even can be as small as one would like if the oscillator is thrown into the black hole sufficiently late. Quantum electrodynamical interaction of blue-shifted cosmic background radiation photons with infalling observers in the same model was addressed in \cite{Burko-97} with similar conclusions. A different approach for finding the interaction of an infalling object with the CH was undertaken by Herman and Hiscock, who considered the increase in internal energy and entropy \cite{Herman-Hiscock-1992}.  (Herman and Hiscock argued that any physical object would be destroyed by such an unbounded increase. See, however, \cite{Ori-97} for a counter argument.) 
The $C^0$ extendibility of the CH was placed on firm mathematical ground for the spherical charged case in \cite{Dafermos-2005} and for the Kerr case in \cite{Dafermos-Luk-2017}.

In this paper we take a similar approach to that taken in  \cite{Burko-Ori-95} in the context of an isolated Kerr black hole perturbed (linearly) by gravitational waves that result from the Price tails that follow the collapse. We model the (test) infalling object by a damped or undamped oscillator, whose extension $d\ll{\cal R}$, where ${\cal R}$ is the typical radius of curvature of the background spacetime between the inner and outer horizons. This assumption allows us to ignore the direct effects of curvature inhomogeneities, and instead treat the object as if it were static in a comoving flat laboratory, subjected to gravitational waves that drive the oscillator. 

The perturbation source we study here is the same source that was considered in \cite{Ori-91,Ori-92,Burko-Ori-95,Burko-97a,Dafermos-2005,Dafermos-Luk-2017,Marolf-Ori-12}, specifically the Price tails that continue falling into the black hole a long time after its formation. Astronomical black holes are believed to be perturbed by additional sources, specifically cosmic background radiation photons  \cite{Burko-97}, baryons, and dark matter \cite{hamilton-2010}. Very recently, a perturbation source of collisionless outgoing and ingoing accretion streams was considered in \cite{hamilton-2017} within a simplifying framework in which spatial gradients are given by a conformally separable solution, possibly beyond the latter's domain of applicability. It was found in \cite{hamilton-2017} that the singularity evolves into a spacelike singularity of possibly the BKL type \cite{BKL}. We do not consider such perturbation sources here. Obviously, our results and conclusions are valid only for the model under consideration.

The main difference between the force driving the oscillator in the spherical charged and Kerr cases is that in the latter case the driving force is oscillatory, and that the frequency of the oscillations, as measured in the infalling laboratory (the object's proper time $\tau$), grows unboundedly on the approach to the CH, whereas in the former case the driving force is monotonic. As the driving force frequency grows, it matches at one time the oscillator's internal resonance frequency. Here, we consider the resonant effect for the first time. 

For specificity, in considering this problem we are loosely motivated by the context of the recent (2014) Hollywood blockbuster movie {\em Interstellar}, in which the protagonist Cooper, strapped in his 
spacecraft, freely-falls towards the inner horizon of a large, rapidly rotating black hole called 
Gargantua. In this work, we are motivated by the ``tidal'' forces experienced by Cooper's 
spacecraft due to the CH singularity, and calculate it in detail for the black hole model used here, and ask whether it could actually 
survive the crossing of the inner horizon\footnote{It should be noted that in {\it Interstellar}, Cooper actually approaches
the outgoing (or ``outflying'') sector of the inner horizon that exhibits a shock-wave like singular 
structure~\cite{Marolf-Ori-12}. However, Kip Thorne (the science advisor for the movie) did not consider that very 
natural, given the expected physical parameters of the orbit that Cooper's spacecraft is on~\cite{KipPvt}. 
In this work, we will study the more physically natural scenario, i.e., Cooper's spacecraft approaching 
the CH (or ``infalling'') sector of the inner horizon. This means that  in this Paper, the inner horizon's singular structure 
develops due to gravitational perturbations that {\em follow} Cooper's spacecraft rather than precede it.} 
by examining the strain on his spacecraft. At a later stage in the movie, Cooper ejects from his spacecraft 
and is shown transitioning into a deeply quantum regime just as he hits Gargantua's inner horizon~\cite{KipPvt}. 
We will not consider any quantum effects in this work: our treatment is based entirely on classical 
general relativity. 

\section{A simple model}

In this section, following the method of \cite{Burko-Ori-95}, we introduce the simple physical model that will allow us to compute some of 
the details of the strain on Cooper's spacecraft through its fall approaching the CH of 
a rapidly rotating Kerr black hole.  

We break this section into two distinct parts: First, we briefly recall the known behavior 
of physical fields, specifically the gravitational field as one approaches the inner horizon 
of an isolated, rotating black hole. This first part is largely a review of earlier work~\cite{Ori1999,BuKhZe2016};  
therefore, we only emphasize the key differences between previous efforts and this current 
work. Second, we describe the physical model we use to understand the effect of these fields on 
a freely-falling, (extended) physical object. Henceforth, $M$ refers to the mass 
of the Kerr black hole and $a$  to its spin parameter. We use standard Boyer--Linquist coordinates 
($t, r, \theta, \varphi$) and also the usual null coordinates $v =  r^* + t$ and $u = {r^*} - t$ (``advanced and retarded times," respectively) 
throughout the paper.  Our equations are in geometrized units of the black 
hole, where distance is measured in units of $GM/c^2$, and time in units of $GM/c^3$, where $G$ is Newton's constant and $c$ is the speed of light in vacuum.

Ori investigated the behavior of physical fields near the CH using a perturbation analysis approach~\cite{Ori-92,Ori1999,Ori2000}, and found the asymptotic behavior of the curvature perturbations in terms of the Weyl scalars  
$\psi _4$ and $\psi_0$ at the early portion of the CH to be
\begin{equation}
\label{eq:psi4}
\psi _4\cong u^{-8}\,(r_--ia\,\cos \theta )^{-4}\sum\limits_{m=-2}^2 
{\,A_m\,_{-2}Y_2^m(\theta ,\phi )\,e^{-im\Omega _-u}}\ ,
\end{equation}
and 
\begin{equation}
\label{eq:psi0}
\psi _0\cong (r-r_-)^{-2}\,v^{-7}\sum\limits_{m=-2}^2 
{\,B_m\,_2Y_2^m(\theta ,\phi )\,e^{im\Omega _-v}}\ .
\end{equation}
Here $r_{-}\equiv M - \sqrt{M^2-a^2}$, $\Omega _-\equiv a/ (2Mr_-)$, $\phi \equiv \varphi 
-\Omega _-t$ is a regular azimuthal coordinate and $_sY_l^m$ are the spin-weighted spherical 
harmonics. $A_m$ and $B_m$ are coefficients that depend on the initial amplitudes of $\psi _4$ 
and $\psi _0$, respectively. In Ref.~\cite{BuKhZe2016} some of us verified these results using a 
direct, numerical approach, i.e., curvature perturbations outside the black hole were directly 
evolved using linearized equations in the background Kerr spacetime of a rotating black hole. 
Care was taken to use a coordinate system that allowed the fields to smoothly evolve right 
through the event horizon and approach the CH along the null direction $u=$ constant, 
$v \rightarrow \infty$. In that work, Ori's results were verified numerically to a high degree of 
accuracy, and new results for the behavior of $\psi_4(u={\rm const},v)$ were found. 

In this current work, we use the same computational setup with one key difference. Instead 
of approaching the CH along a null geodesic, we approach it along a time-like one. 
This allows us to directly study the effect of physical fields on a freely-falling object as 
it approaches the CH of a rotating black hole. 

In the context of a rotating black hole, it was shown in Ref.~\cite{Ori1999} that the most divergent 
components of the Riemann curvature tensor are proportional to $\psi_0$, as given in (\ref{eq:psi0}).  
Ori predicted that the ``tidal'' forces on an infalling object would behave as the real part of
\begin{equation}
\tau ^{-2}\,\left[ {\ln \left( {-\tau / M} \right)} \right]^{-
7}\,\sum\limits_{m=1,2} {C_m\,_2Y_2^m}(\theta _0,\phi _0)\,\,e^{-
imp\ln \left( {-\tau / M} \right)}\,
\label{force}
\end{equation}
where $\tau$ is the proper time associated to the falling object chosen such that $\tau = 0$ 
is when the object hits the CH, and $\theta _0,\phi _0$ are the 
angular positions at which the object hits the CH. Here, $p\equiv a\,(M^2-a^2)^{-1/ 2}$. This expression for the tidal forces is valid 
when the object is close to the inner horizon, i.e., when $\tau$ has a sufficiently small negative 
value.

Ori argued that the tidal deformations described by (\ref{force}) would stay finite \cite{Ori-92}, thus making the CH 
singularity a  weak one in Tipler's classification~\cite{Tipler}. (See also \cite{Ori-2000}.) It was also noted in Ref.~\cite{Ori-92} that the 
object would experience infinitely many oscillations, with ever increasing frequency. We verify  below all of these expectations by performing direct numerical computations. 

To study the effect of the tidal forces on an infalling extended object, we consider 
a basic oscillator as a physical toy model for an extended object. The oscillator is parametrized by mass $\mu$, spring 
constant $k$, and damping coefficient $b$. Our mathematical model for the infalling object 
will be given by
\begin{equation}
\mu {\ddot x} + b {\dot x} + k x = F(\tau)\ , 
\label{oscillator}
\end{equation}
where $x(\tau)$ is proportional to the strain experienced by the object. (At equilibrium length $x=0$.) The driving 
force $F(\tau)$ is  chosen proportional to the expression~(\ref{force}). In the 
next section we show that our numerical results for $\psi_0$ corroborate this expression for $F(\tau)$,  
and find $x(\tau)$ throughout the fall.  We align the model object so that it responds to the driving force in full magnitude, i.e., the infalling spacecraft uses its engines to align itself so that it does not rotate. 

\section{Computational results}

We begin this section with a brief description of our numerical setup. As pointed out 
in the previous section, we take an approach similar to the one described in detail 
in Ref.~\cite{BuKhZe2016}. In particular, we numerically solve the Teukolsky equation 
for $\psi_0$ and  $\psi_4$ in specifically designed compactified, ingoing-Kerr coordinates 
that allow us to ``penetrate'' the horizons and also provide us with great computational 
efficiency. The initial field configuration is a Gaussian profile located outside the 
horizon, centered at $r = 8M$ and of width $M/10$. The angular distribution of the initial 
field corresponds to $\ell = m = 2$. Other numerical parameters and details of the 
numerical scheme utilized may be found in Ref.~\cite{BuKhZe2016}.
As the computation evolves the fields are sampled along a time-like geodesic. 
For the results depicted below, equatorial-geodesics with $E/\mu = 1$, $L/\mu = 2M$ were used; 
however, the main features of our results appear to be applicable to other time-like 
geodesics. In Fig.~1 we show a plot for the late portion of the time-like geodesics we 
used for two Kerr black holes with spin parameters $a/M = 0.8$ and $0.9165$.\footnote{In {\it Interstellar}, Gargantua's mass is $M_{\rm Gar.}\sim 1-2\times 10^8\, M_{\odot}$ and its spin parameter is $(a/M)_{\rm Gar.}\sim 1-\varepsilon$, where $\varepsilon\sim 1.3\times 10^{-14}
$ \cite{KipPvt}. Such closeness to extremality is difficult to simulate numerically, so that we  simulate in practice lower values for $a$, although we still solve for a fast spinning Kerr black hole.}

\begin{figure}
\includegraphics[width=7.5cm]{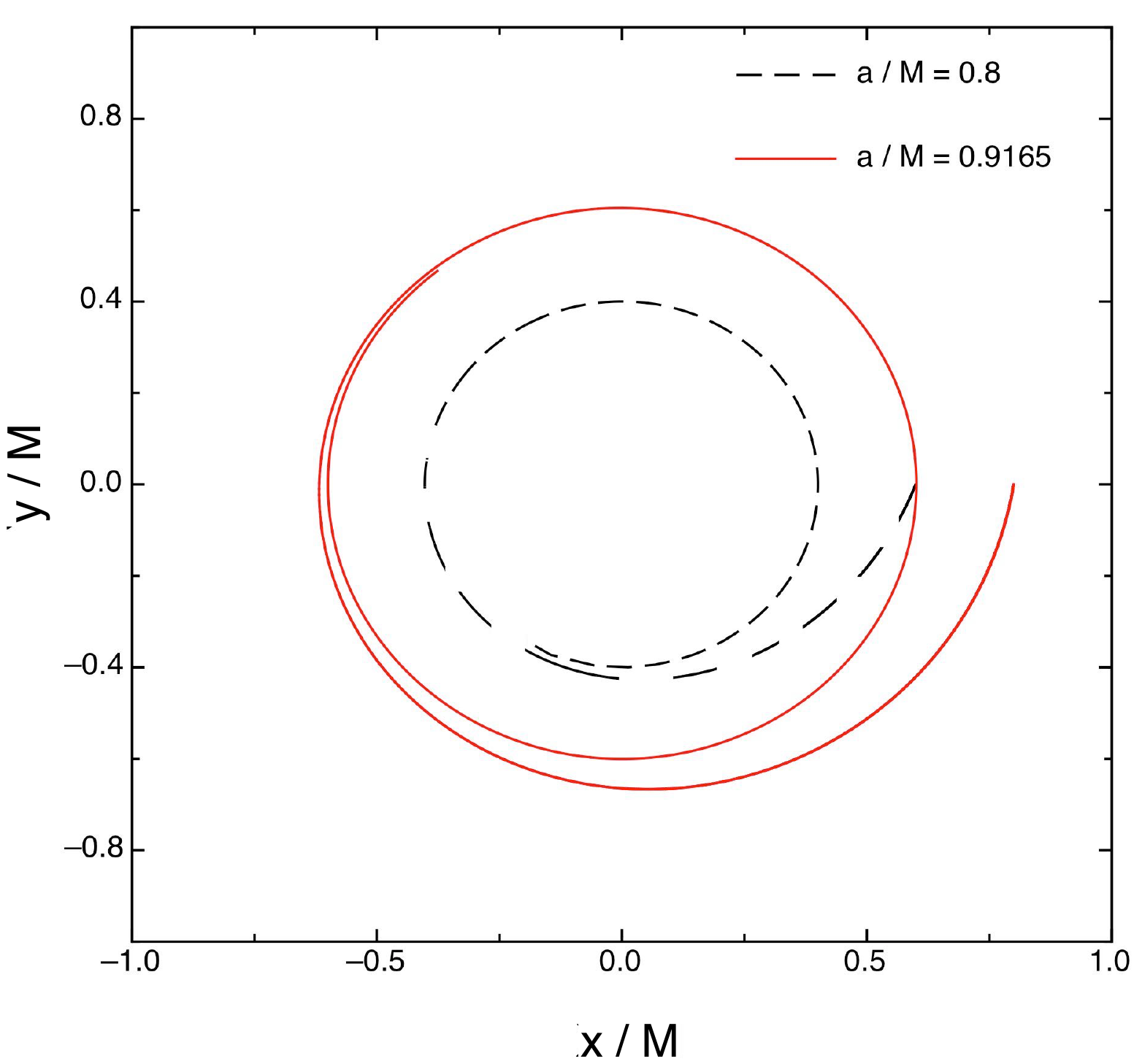}
\caption{The $E/\mu = 1$, $L/\mu = 2M$, equatorial, time-like geodesics for two black holes 
with spin rates $a/M=0.8$ or $0.9165$. The inner horizons are, correspondingly, at $r_-=0.4M$ or $0.6M$. We depict the late stage portion of these geodesics.}
\end{figure}

\subsection{Oscillatory singularity}

In Figs.~2,~3 we show the results of our numerical simulations, alongside the results 
of perturbation analysis \cite{Ori1999} for the time-like geodesics shown in Fig.~1. There is clear agreement between the two results, 
especially at late-times; note that the overall amplitude and phase were adjusted to 
obtain this match, and that the agreement is maintained over more than 50 orders of magnitude. At early times, we do not expect to see a close match between the 
two results because in Ref.~\cite{Ori1999} multiple near-horizon approximations were used in the analysis, whereas our results capture also subdominant modes. 

\begin{figure}
\includegraphics[width=7.5cm]{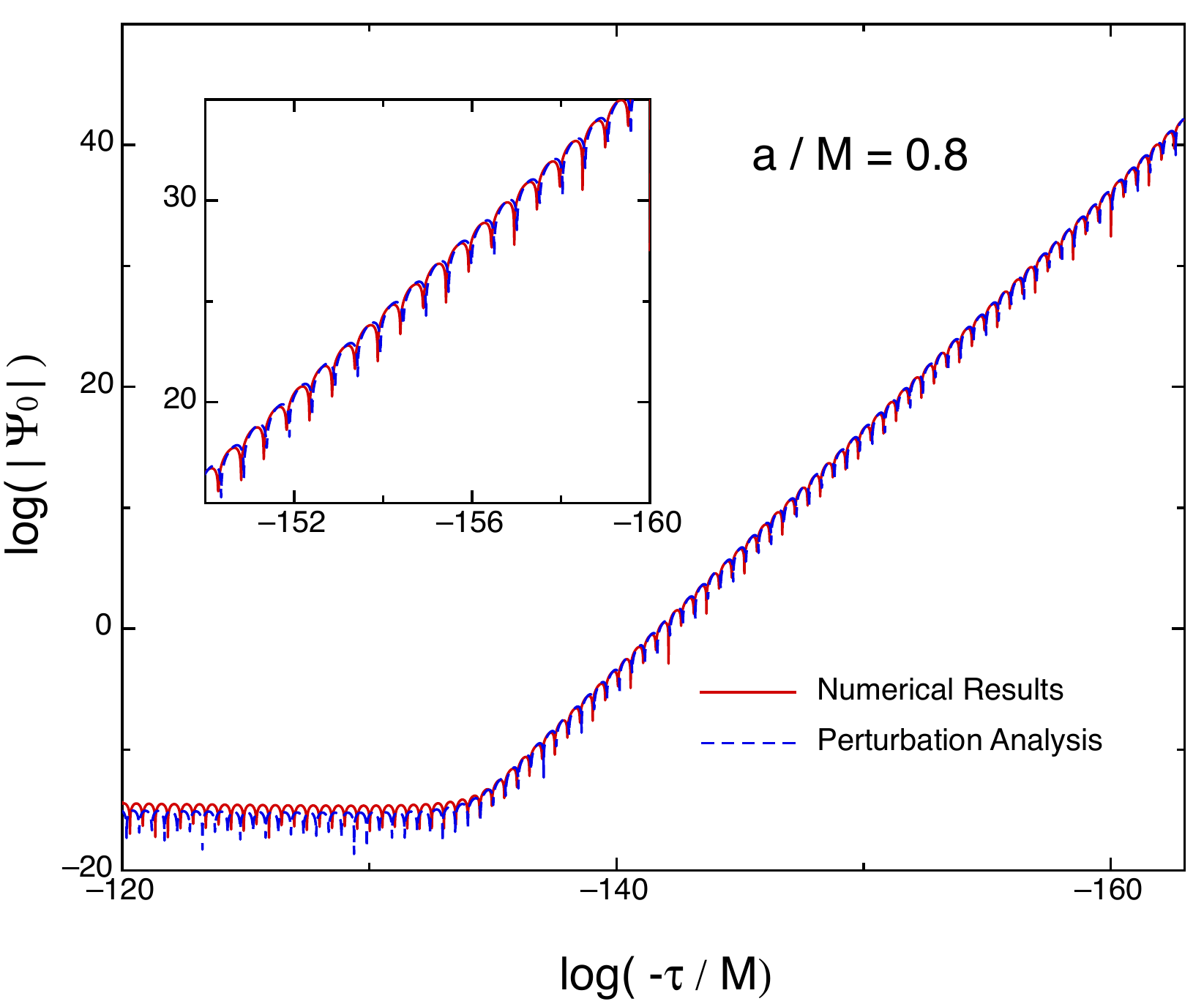}
\caption{The curvature scalar $\psi_0$ as a function of the proper-time of a 
time-like observer falling into a Kerr black hole with $a/M = 0.8$. The logarithms are base 10. Our numerical results are depicted by a solid curve, and the perturbation analysis prediction of Ref.~\cite{Ori1999} are shown with a dashed curve.} \label{fig2}
\end{figure}

It is clear that at late times, i.e., in close proximity to the CH, 
curvature grows with an inverse-square power of $\tau$. (Recall that $\tau < 0$ and 
approaches zero). This unbounded growth of curvature is the physical manifestation of the failure of the manifold to have 
square-integrable connections in any neighborhood of the CH. 
Naively, one may interpret that growth rate as a divergence that would destroy any 
object that is unfortunate enough to approach it. We address this issue in subsection \ref{subsection:extended}. In addition, notably the frequency of oscillation monotonically increases with proper time $\tau$ upon approach to the CH. In fact, 
the instantaneous frequency can easily be computed to be $\omega(\tau) = 2 p / (-\tau)$. 
Thus, the infalling object experiences ever stronger tidal forces and with faster oscillatory 
cycles as it approaches the CH. 

\begin{figure}
\includegraphics[width=7.5cm]{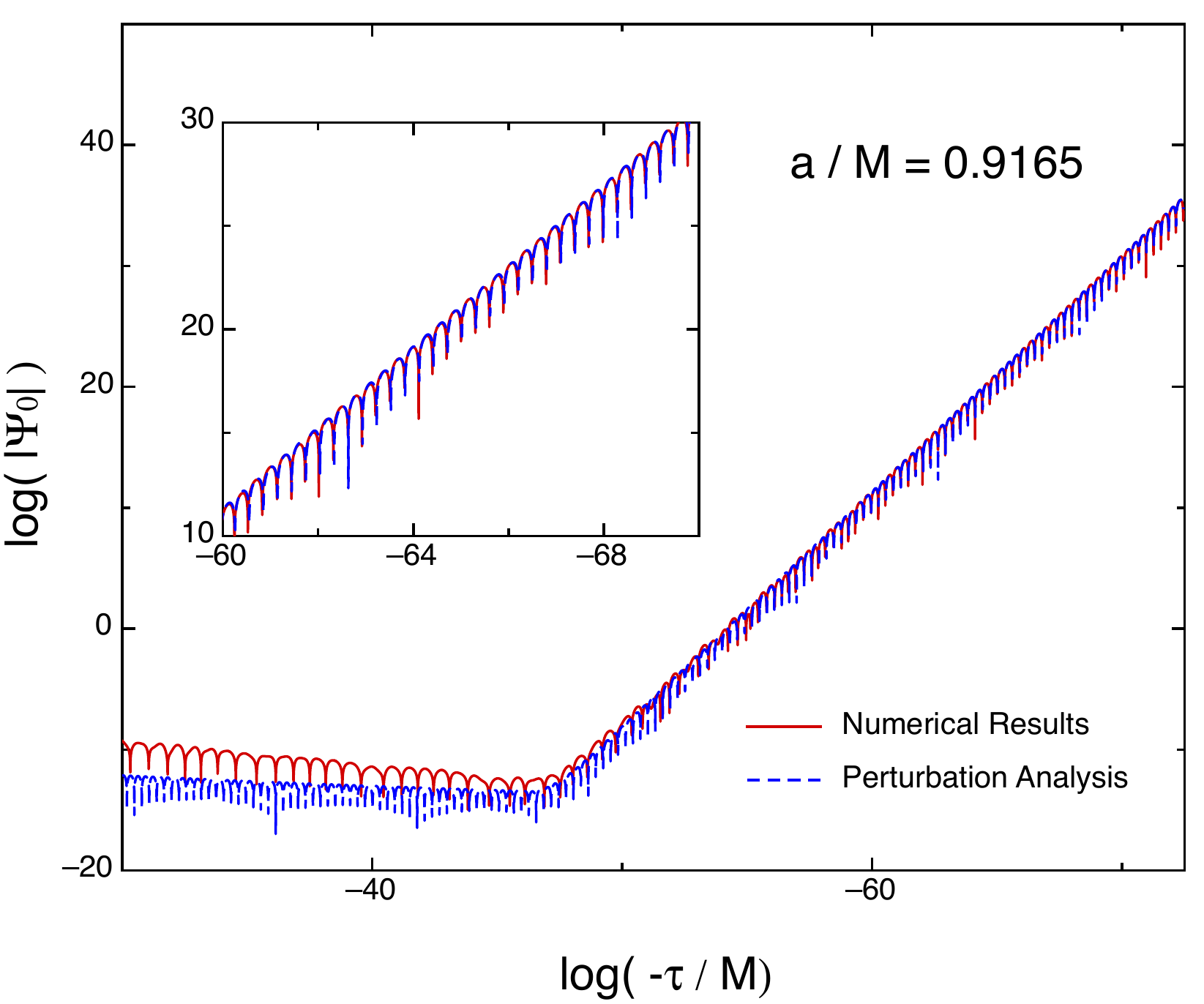}
\caption{Same as Fig.~\ref{fig2} for $a/M = 0.9165$.
} 
\end{figure}

\subsection{Extended physical object}\label{subsection:extended}

We use the simple oscillator model Eqn.~(\ref{oscillator}) for understanding the effects of the tidal forces~(\ref{force}) on a material object. 

First, consider the form of the forcing function $F(\tau)$ in Eqn.~(\ref{oscillator}).  This expression should be proportional to the tidal force given in the formula~(\ref{force}).  For simplicity, we focus on the $\ell=2,m=2$ harmonic, which is the dominant harmonic near the black hole's equator. We are concerned with the real part of the expression~(\ref{oscillator}), so the $e^{-i m p \log(-\tau/M)}$ factor can be simplified to $\cos{\left[ m p \ln(-\tau / M)\right]}$. The $\ell=2,m = 2$ harmonic $Y^2_2$ is proportional to a constant, so we are left with the driving force 
\begin{equation}
\label{FofTau}
F(\tau) = F_0\,\tau ^{-2}\,\left[ {\ln \left( {-\tau / M} \right)} \right]^{-7}\,\cos{\left[2 p \ln(-\tau / M)\right]}
\end{equation}
where $F_0$ is some constant that depends on the details of the actual physical origin of the black hole perturbation and the time that has elapsed since it formed or since its mass has increased until the object crosses its event horizon. The calculation of $F_0$ is outside the scope of this work, so we will treat it as an overall scaling constant for the strain. In practice, our results for the strain are the strain per $F_0$.

In geometrized units the parameters of the oscillator (\ref{oscillator}), $\mu$, $b$, and $k$,  turn out to be many orders of magnitude apart, for any realistic oscillating matter falling into a supermassive black hole like Gargantua. Recall that the parameters $\mu$, $b$, and $k$ have fixed values  associated to the properties of the infalling object. Furthermore, $F(\tau)$ oscillates rapidly as $\tau \to 0^-$.  The equation is stiff and difficult to simulate. 

We can facilitate the simulation by replacing $\tau$ with a new time variable $s$, where $s \equiv - M \ln (-\tau/M)$, (``logarithmic time") for which the object hits the CH as $ s \to \infty$.  The use of $s$ as the time variable requires some alteration to the form of the Eqn.~(\ref{oscillator}), since the strain $x$ will now be expressed as a function of $s$ rather than $\tau$.  We can rewrite Eqn.~(\ref{oscillator}) as 
\begin{equation}
\mu {x''} + \left(\frac{\mu}{M} + b e^{-s/M}\right) x' + e^{-2s/M} k x = -\frac{{F_{0} M^5}\cos{\frac{2 p s}{M}} }{ s^{7}}
\label{oscillator_s}
\end{equation}
where $x''$ and $x'$ refer to the derivatives of $x$ with respect to logarithmic time $s$. 

\subsubsection{Analytical approximation}
In addition to being easier to simulate, this form of the oscillator Eqn.~(\ref{oscillator_s}) makes it easier to set some qualitative expectations of what may happen as the falling object approaches the CH. The forcing function on the right-hand side maintains a constant frequency of $2p / M$ when $s$ is the time coordinate.  However, the forcing function decreases in magnitude rapidly. At large values of $s$ the oscillator Eqn.~(\ref{oscillator_s}) can be reduced to the trivial form $x''(s) + x'(s) \cong 0$. This implies that the ``strain'' $x(s)$ has an exponentially decaying solution (to a constant asymptotic value) as $s \to \infty$. 

However, this exponentially decaying strain does not rule out the possibility of an interesting phenomenon at a somewhat earlier time: Specifically, note that while the forcing frequency is a constant $\omega(s) = 2 p/M$, the effective natural frequency of the object is not constant. Recall that the $x$ term of a standard drived oscillator equation (\ref{oscillator}) is proportional to natural frequency squared ($k = \mu \omega_0^2$). In our altered oscillator equation (\ref{oscillator_s}), the $x(s)$ term  is instead proportional to $e^{-2s/M} k$.  That is, $\mu \tilde{\omega}_0^2 \equiv e^{-2s/M} k$, where $\tilde{\omega_0}$ is the effective natural frequency, and 
\begin{equation}
\label{w0}
\tilde{\omega}_0 = e^{-s/M} \sqrt{k / \mu} \, .
\end{equation}
This result suggests that 
the natural frequency effectively decreases with logarithmic time $s$. At some point during the object's fall, its effective natural frequency $\tilde{\omega}_0$ may be equal to the forcing frequency $\omega(s)$. This occurs at logarithmic time
\begin{equation}
\label{sres}
\tilde{s} = -M\ln{\left[ \frac{2p}{M}\sqrt{\frac{\mu}{k}}\right]}\, .
\end{equation}
We expect that as $s \simeq \tilde{s}$, a resonance-like effect may occur. Such an effect may result in a momentary increase in the strain on the falling observer. We direct our attention to these expectations in our numerical results that appear below.

\subsubsection{Numerical results} 
 
Our numerical simulations consist of solving Eqn.~(\ref{oscillator_s}) for different input parameters. We primarily used MATLAB R2017a's built-in stiff ODE solvers.  Care must be taken when solving ODEs like Eqn.~(\ref{oscillator_s}):  Simulations which are set up incorrectly can yield very different results when different methods are used, or when simulating forward or backward in time.  To ensure accuracy, we present results which were nearly identical for all ODE solvers tested.  As a check, we simulated both forward and backward in time, to ensure convergence in the region of interest.  Finally, we checked our results against results from the \texttt{ParametricNDSolve} function in Mathematica 11.  

The four MATLAB ODE solvers tested were \texttt{ode15s}, \texttt{ode23s}, \texttt{ode23t}, and \texttt{ode23tb}.\footnote{
\texttt{ode15s} is a variable-step, variable-order solver based on the numerical differentiation formulas of orders 1 to 5. \texttt{ode23t} is an implementation of the trapezoidal rule using a ÒfreeÓ interpolant. 
\texttt{ode23s} is based on a single-step modified Rosenbrock formula of order 2.
\texttt{ode23tb} is an implicit Runge-Kutta formula with a trapezoidal rule step as its first stage and a backward differentiation formula of order 2 as its second stage. See \cite{matlab-ode}.}  These solvers do not have a fixed step-size.  Instead, they use numerical derivatives to estimate an appropriate step-size adaptively.  When MATLAB was allowed to choose the step-size, the different solvers did not give converging results.  However, if the solvers' \texttt{MaxStep} option is set to a sufficiently low value, results either converge, or the solvers fail entirely and do not run.  We found that  \texttt{ode23s} and \texttt{ode23tb} converged at $\texttt{MaxStep} = 10^{-3}M$ or lower, and \texttt{ode15s} and \texttt{ode23t} failed entirely (see Fig.~\ref{steelosc_vs}). 
   
We also checked whether \texttt{ode23s} gave the same result when evolved forward in time (in the $+s$ direction) and backward in time (the $-s$ direction).  The $-s$ evolution tends to diverge from the $+s$ evolution at low $s$, but is well-behaved for lower values $s$ when \texttt{MaxStep} is further reduced. 
     
\begin{figure}
\includegraphics[width=8.5cm]{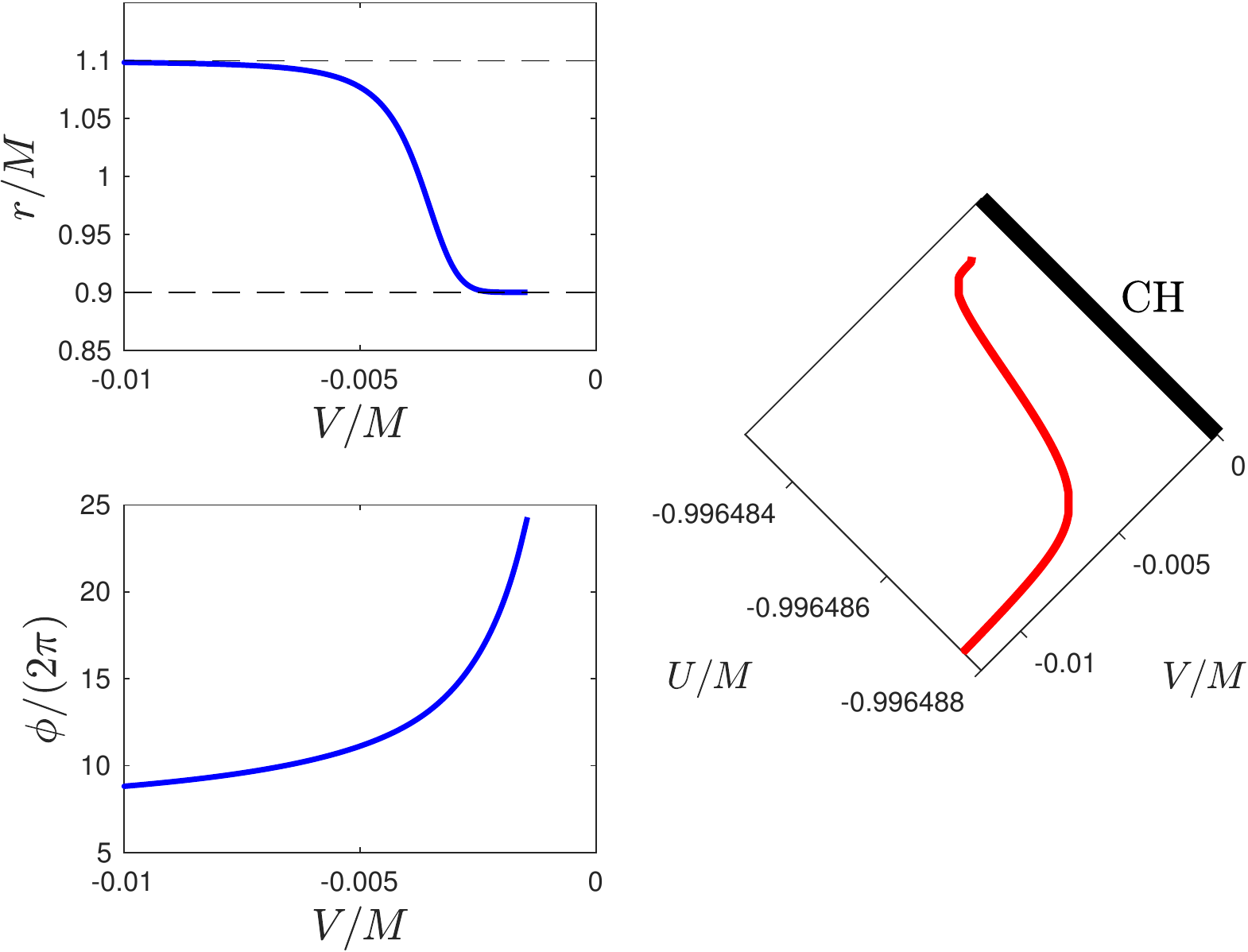}
\caption{The path of an object with $E/\mu=1$ and $L/\mu=4M$ falling toward the CH of a Kerr black hole with $M=10^6\,M_{\odot}$ and $a/M=0.995$ ($p=10$) on the equatorial plane. The CH is at Kruskalized advanced time $V=0$. Top left panel: $r$ as a function of $V$. The event and inner horizons are shown in dashed lines. 
Bottom left panel: $\phi$ as a function of $V$. Right panel: The Penrose diagram.  It is as yet unknown whether there is a continuation of the spacetime manifold beyond the CH (thick line). Notice, that Kruskalized retarded time $U$ has been stretched in the Penrose diagram to show the curvature in the world line. Because of the scale of the diagram the event horizon (at $U/M=-1$) or future null infinity ${\mathscr I}^+$ are not shown on the Penrose diagram.} 
\label{PD}
\end{figure}

\begin{figure}
\includegraphics[width=8.5cm]{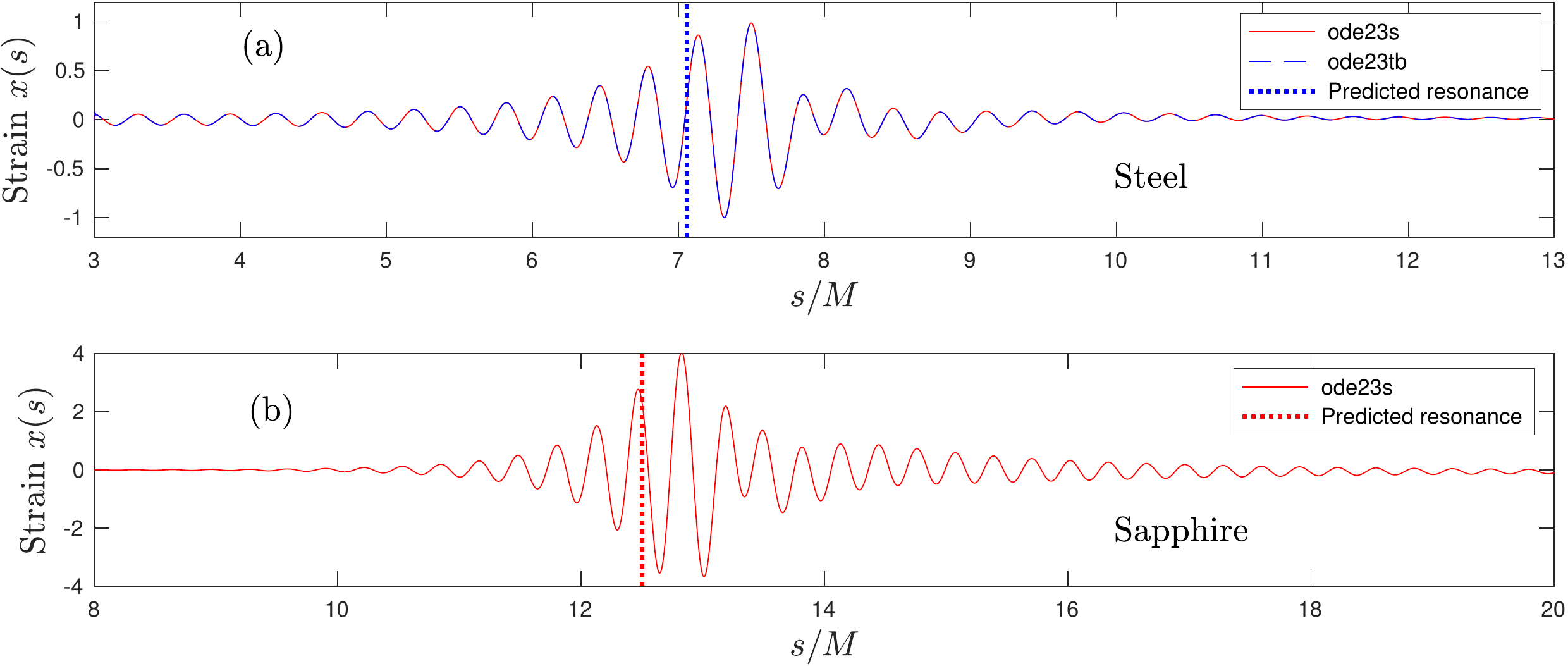}
\caption{The normalized strain of a $1\,{\rm m}^3$ cube of 1020 steel (top panel, (a)) and on a $1\,{\rm m}^2\times 0.01\,{\rm m}$ thick plate of single-crystal sapphire (bottom panel, (b)), as they approach the CH of a Kerr black hole as functions of logarithmic time $s$. In the top panel (a) the simulation is done using both MATLAB's \texttt{ode23s} and  \texttt{ode23tb} methods  with \texttt{MaxStep} set to $10^{-3}M$. In the bottom panel (b) the simulation is done using MATLAB's \texttt{ode23s} method with \texttt{MaxStep} set to $10^{-3}M$. In both panels the expected time of resonance is indicated with dotted lines. The results are normalized so that the peak strain of the steel cube is unity. Note, that the strain $x$ is shown here (and also in Figs.~\ref{steelosc_undamped} and \ref{steelosc_undamped_in_t}) in geometrized units.}
\label{steelosc_vs}
\end{figure}

For specificity, we choose a fast spinning Kerr black hole, with $a/M=0.995$ ($p=10$), and choose the time-like geodesic to have $E/\mu=1$ and $L/\mu=4M$. The shape of this geodesic is shown on the Penrose diagram in  Fig.~\ref{PD}. Our choice of the input material parameters ($\mu$, $b$, $k$) is much more open.  The vibration of the infalling object sweeps across an infinite range of frequencies of the incoming gravitational waves: It is easy to design inputs that meet the resonance condition at some point during the infall.  However, not all such input parameters are physically reasonable, and it can be difficult to intuit what physically reasonable parameters look like when working in the natural units of a black hole.  We would like to demonstrate that the resonance conditions can be met by realistic objects, and are not just a mathematical curiosity that may only be relevant to unrealistic matter.  To that end, we present results where $\mu$, $b$, $k$ are set to values corresponding to actual materials. 

First, we choose the parameters to roughly correspond to a $1\, m^3$ steel block, whose vibrations are damped at 1/4 of critical damping ($\zeta = \frac{1}{4}$). Here, the damping ratio $\zeta:=b/(2\sqrt{\mu k})$. These parameters are listed in Table \ref{tab:mbk_steel}.  The middle column of Table \ref{tab:mbk_steel}  shows the parameters in MKS units, while its last column shows the same values, as converted into the geometrized units of a black hole with $M = 10^6 M_{\odot} = 2 \times 10^{36} \,{\rm kg}$.

\begin{table}[h!]
  \begin{center}
    \caption{Parameters for a $1\,m^3$ of damped 1020 steel, for $10^6\,M_{\odot}$ black hole. $^\dagger$: Calculated using $k = YA / L$, where $Y$ is Young's modulus and $A$ and $L$ are the area and length of the cube, respectively. Data are from Ref.~\cite{steel}.
}
    \label{tab:mbk_steel}
    \begin{tabular}{l|c|r} 
      \textbf{Parameter} & \textbf{MKS values} & \textbf{Geometrized values}\\
      \hline
      $\mu$ & 7870 kg & $3.935 \times 10^{-33}$\\
      $b\,\,(\zeta = \frac{1}{4})$ & $1.913 \times 10^7$ N/(m/s) & $4.738 \times 10^{-29}$\\
      $k$ & 186 GPa $\cdot$ m  $^\dagger$ & $2.282 \times 10^{-24}$\\
      \hline
    \end{tabular}
	     \end{center} 
\end{table}      
\noindent Notice that in geometrized units of a supermassive black hole, the parameters of an ordinary object may be quite small, as is the case for the steel cube.  We estimate that in order to see resonance, we need the expected resonance logarithmic time $\tilde{s}$ to attain a positive value (see Eqn.~(\ref{sres})). Therefore, the size of the parameters is less relevant than the ratio of $\mu$, $k$, and $p$.  For a high-spin black hole (e.g. $p = 10$ or $a/M = 0.995$), $k$ must be several orders of magnitude higher than $\mu$ in order for $\tilde{s}$ to be positive. This is the case for the steel cube parameters, for which $\tilde{s} = 7.06M$. Although we could choose $\mu$ and $k$ to be larger numbers, there is little to be gained in terms of either mathematical elegance or realism.  Note, however, that to make a precise prediction we would be required to have at least an order-of-magnitude estimate for $F_0$ in Eqn~(\ref{oscillator_s}).  

The object's oscillation is directly proportional to $F_0$, but we are agnostic here to the magnitude of $F_0$, because there are many possible sources for the perturbations of the black hole, e.g., the remnant of fields associated with the collapse process, accreted matter from a surrounding disk, perturbations from an orbiting spacecraft, photons coming from the cosmic background radiation, etc. 
Therefore, the results of our simulation of Eqn.~(\ref{oscillator_s}) are normalized so that the peak oscillation amplitude is unity.  The normalization factor is $7.4166\times 10^{-25}$. The smallness of the normalization factor will become pivotal for our conclusions below. These results are shown in Fig.~\ref{steelosc_vs}. Note, however, that for an object that falls into the black hole a long time after the latter has formed (from a collapse process or a long time after its mass has increased otherwise) and has been unperturbed otherwise, the effective magnitude of $F_0$ is very small due to the exponential decay of quasi-normal modes followed by a Price power-law tail. We have not factored out such a decrease in the magnitude of $F_0$ here. For an old black hole (i.e., a black hole which is unperturbed since formation), such as Gargantua, the normalization factor will therefore be much smaller than the one we are using here. The magnitude of our normalization factor is perhaps more relevant for a young black hole, i.e.,  for an object that falls into the black hole only a short time after the latter has formed. In this sense our results are upper bounds on the response of the infalling object. 

Second, for comparison we change the parameters for an object made of a different material and having a different shape.  In Table \ref{tab:mbk_sapphire}, we consider a $1\,{\rm m}^2\times1\,{\rm cm}$  plate of single-crystal sapphire (Al$_2$O$_3$), for which the interaction is across the thin side.  This material can be used in aerospace viewing windows, so we could imagine this plate as a window in a spacecraft.  As with the steel cube, we are able to predict the  resonance time of $\tilde{s} = 12.5M$ with reasonable accuracy.  The results for the sapphire plate window are also shown in Fig.~\ref{steelosc_vs}.  It is apparent that the steel cube and sapphire plate window behave qualitatively similarly, and that both resonate in a predictable but distinct times.

\begin{table}[h!]
  \begin{center}
    \caption{The parameters for a $1\,{\rm m}^2$ by $0.01\,{\rm m}$ thick plate of single-crystal sapphire, damped, for $10^6\,M_{\odot}$ black hole.  $^\dagger$: Calculated using $k = YA / L$, where $Y$ is the Young's modulus and $A$ and $L$ are the area and thickness of the plate, respectively.
Data are from Ref.~\cite{sapphire}.}
    \label{tab:mbk_sapphire}
    \begin{tabular}{l|c|r} 
      \textbf{Parameter} & \textbf{MKS values} & \textbf{Geometrized values}\\
      \hline
      $\mu$ & 39.7 kg & $1.985 \times 10^{-35}$\\
      $b\,\,(\zeta = \frac{1}{4})$ & $2.160 \times 10^7$ N/(m/s) & $5.349 \times 10^{-29}$\\
      $k$ & 470 GPa $\cdot$ m  $^\dagger$ & $5.767 \times 10^{-22}$\\
      \hline
    \end{tabular}
	     \end{center} 
\end{table}      

\begin{figure}
\includegraphics[width=8.5cm]{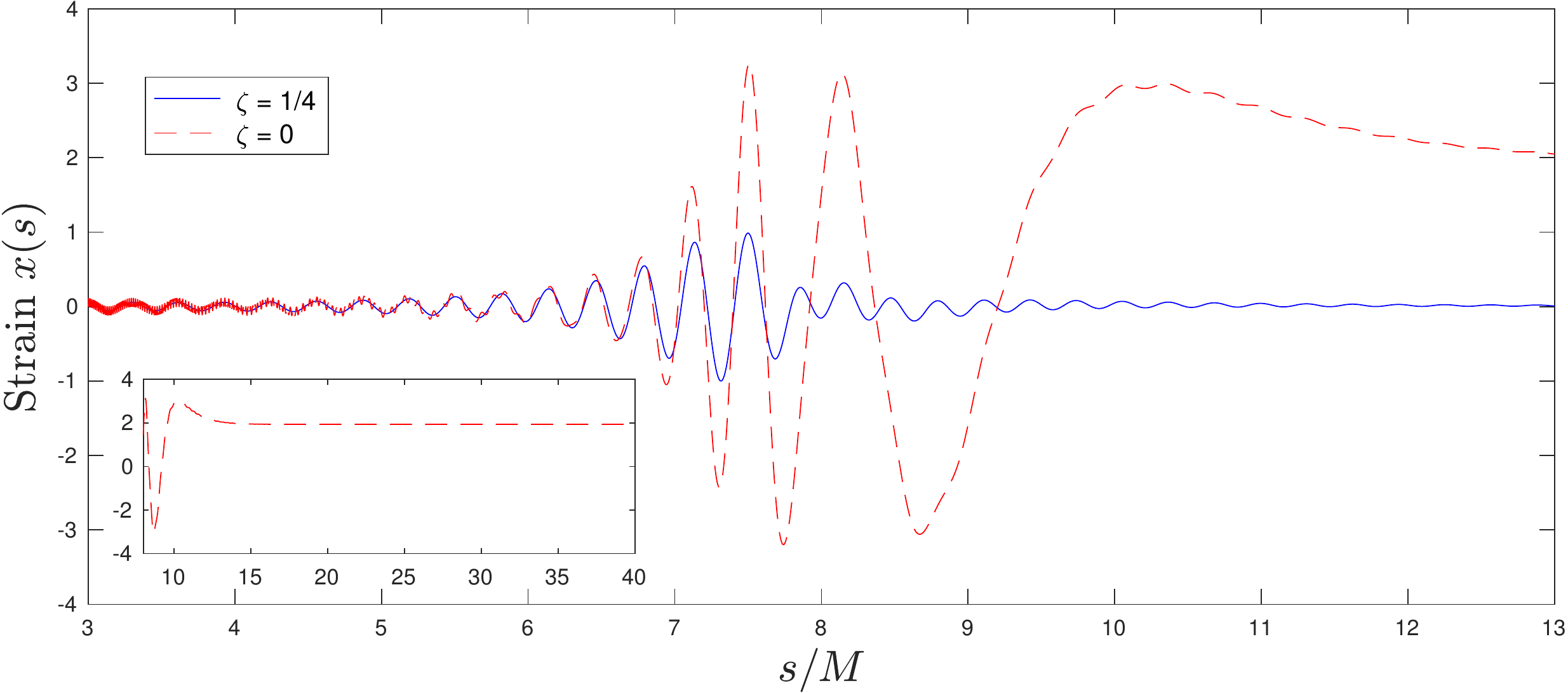}
\caption{The damped ($\zeta=1/4$, solid) and the undamped ($\zeta=0$, dashed) strains on the steel cube as functions of logarithmic time $s$. The inset shows the undamped case at later values of $s$ than is shown in the main figure. The scaling is chosen so that maximum strain of the damped case has unity value. The maximum strain of the undamped case on this scale is 3.2585, and the undamped oscillators settles to a permanent strain of 0.5935 of its maximum strain. }
\label{steelosc_undamped}
\end{figure}

We may also wish to consider the role of damping in this oscillation.  Removing the damping from the steel cube (the model considered in \cite{Burko-Ori-95}), we observe the expected higher amplitude of oscillation, as shown in Fig.~\ref{steelosc_undamped}.  There are also some differences in the frequency composition, which are most obvious at times later than the resonance.

\begin{figure}
\includegraphics[width=8.5cm]{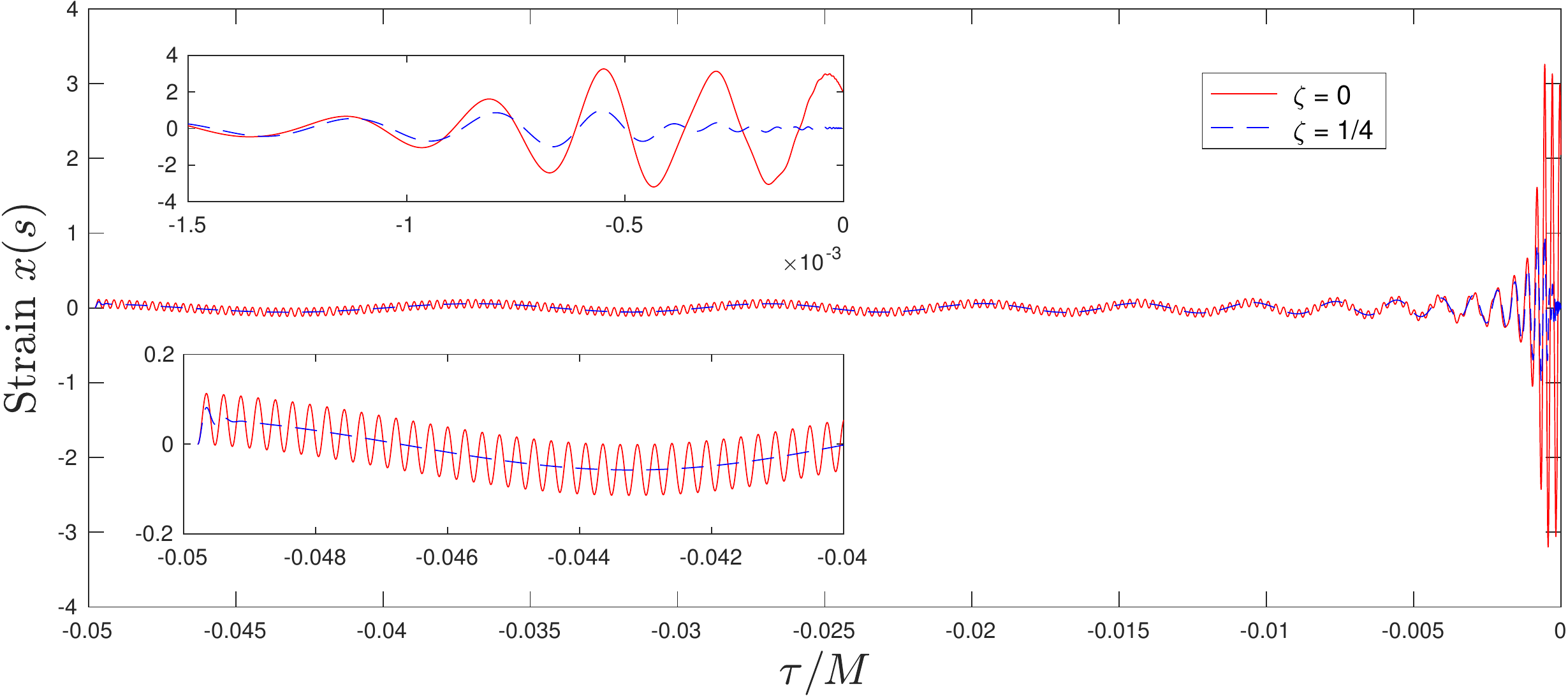}
\caption{The damped ($\zeta=1/4$, dashed) and the undamped ($\zeta=0$, solid) strains on the steel cube as functions of proper time $\tau$. The insets show zoom in for early and late proper times. The scaling is chosen so that maximum strain of the damped case has unity value. }
\label{steelosc_undamped_in_t}
\end{figure}

To see the early-time differences between the damped and undamped oscillations, it is best to view the data in proper time $\tau$ instead of the logarithmic time coordinate $s$.  Fig.~\ref{steelosc_undamped_in_t} is the same as Fig.~\ref{steelosc_undamped}, but shown in $\tau$.  Additionally, Fig.~\ref{steelosc_undamped_in_t} contains an inset showing the detail of the damped versus undamped oscillation at low $\tau$.  We see that the undamped oscillations include undulations which appear to have constant period in $\tau$, not $s$.  These oscilations can be measured to have a period of about $2.6 \times 10^{-4}$ in $\tau$, which implies an angular frequency of $\omega(\tau) \approx 2.4 \times 10^4$.  We note that in the time domain, our steel cube has an analytic natural frequency of $\omega_0(\tau) = \sqrt{\frac{k}{\mu}} = 2.4082 \times 10^4$, similar to the measured frequency of these undulations.  It appears that in the absence of damping, an infalling object may vibrate near its natural frequency even before resonance occurs.  The resonance will occur when the frequency of the driving force  and this natural frequency approximately coincide.  

\section{Concluding remarks}

We have shown that the total integrated deformation for a simple model of a realistic physical object approaching the CH of an isolated spinning black hole which is perturbed by the Price tails of radiation that follow the collapse process is bounded, and that the maximal deformation may be obtained well before hitting the CH, when the incoming gravitational waves are in resonance with the object's natural frequency. The actual deformation depends on the parameter $F_0$, which in turn depends on the perturbing field of the black hole and the age of the black hole at the moment the infalling object crosses the event horizon. One should bear in mind that in addition to the deformation itself, any physical object would also undergo successive oscillations, which may undermine its integrity and strength if they are violent enough. In practice, the survivability of the material depends on the interplay of the material's relaxation time with the strain rate and the oscillations frequency. 
However, we predict that these deformations and oscillations are quite small. 

Our interaction model is very simplistic. A realistic object would experience tidal deformations along all three axes, so that the object is successively stretched along one direction and compressed in the other directions. Also, because of the effect of dragging of inertial frames, such an object will generally rotate, so that the direction along which the tidal forces act will change. One approach to address this effect would be to solve for the Jacobi fields along the object's worldline to find the time behavior of the volume element. We do believe, however, that our very simple physical model captures the essential part of the tidal deformation of a realistic object approaching the CH of a black hole within the model used.


This paper does not address the deformation suffered by objects that approach the outgoing leg of the black hole's inner horizon, where one expects to find the Marolf-Ori ``outflying" singularity \cite{Marolf-Ori-12}. This paper also does not address other perturbation sources such as baryons and dark matter from accretion processes, and irradiating photons from the cosmic background radiation, specifically those associated with an asymptotically de Sitter universe. Recently there has been renewed interest in the SCC in the context of black holes that are immersed in de Sitter spacetime \cite{Cardoso-2018,Dias-2018}. Detailed interaction model for the Kerr -- de Sitter case is awaiting further investigation. 

Finally, we have addressed the interactions until the moment that the infalling object hits the CH. It is as yet an open question whether, within the model studied here, there is a classical continuation of the spacetime manifold beyond the CH, perhaps following a short transitional regime where quantum gravity effects are important, and whether physical objects could survive  this transition. 
  
\section*{Acknowledgements} 
We thank Amos Ori for discussions at an early stage of this project, and for valuable comments made on an earlier draft. We thank Kip Thorne, Jay Wang, and Scott Field for discussions. C.M.~acknowledges research support from the University of Massachusetts Dartmouth Graduate School. G.K.~acknowledges research support from NSF Grants No.~PHY--1701284, and from the U.S.~Air Force agreement No.~10--RI--CRADA--09.

\end{document}